\documentclass[pre,10.5pt]{revtex4}

\newcommand{\BF}[1]{\mbox{\boldmath$#1$}}
\def\al{\alpha}

\def\be{\begin{equation}}
\def\ee{\end{equation}}
\def\bea{\begin{eqnarray}}

\def\eea{\end{eqnarray}}
\def\la{\label}
\def\bsea{\begin{subeqnarray}}
\def\esea{\end{subeqnarray}}
\def\u{\underline}

\begin{document}
\title{Mode coupling theory in
the FDR-preserving field theory of interacting Brownian particles}
\author{Bongsoo Kim${}^ 1$ and Kyozi Kawasaki${}^ 2$}
\affiliation{${}^ 1$ Department of Physics, Changwon National University,
Changwon 641-773, Korea\\
${}^ 2$ Electronics Research Laboratory,
Fukuoka Institute Technology, Fukuoka, Japan}
\date{\today}
\begin{abstract}
We develop a renormalized perturbation theory for the
dynamics of interacting Brownian particles,
which preserves the fluctuation-dissipation relation
 order by order.
We then show that the resulting one-loop theory gives a closed
equation for the density correlation function,
which is identical with that in the standard mode coupling theory.\\
PACS 64.70.Pf
\end{abstract}
\maketitle

\setcounter{equation}{0}
As a first-principle approach, the mode coupling theory (MCT) \cite{goetze,das,reich}
has not only enjoyed considerable success in explaining 
the slowing down of the weakly supercooled liquids,
but also had enormous impact on the area by 
stimulating further experiments, simulations,
and other theoretical developments.
However, the foundation of the theory needs much to be desired
since the theory is fraught with uncontrolled approximations.
It is helpful if one can develop systematic field-theoretical treatment for
model system containing smallness parameter.
The earlier field-theoretic formulations \cite{kkft,dasmazenko} of MCT
are found to be incompatible with the fluctuation-dissipation relation (FDR)
\cite{miya,abl}. It was thus an urgent task to develop a
consistent field-theoretic formulation capable of describing
the dynamics of glass-forming liquids,  for which
a systematic perturbation expansion preserving the FDR is possible.
Quite recently, Andreanov, Biroli, and Lefevre (ABL)\cite{abl}
 provided a remarkable insight into this problem by
focusing on the symmetry properties of the action integral under
time-reversal (TR). In particular, ABL  has identified the transformation of fields
 under TR, which leaves the action invariant.
The FDR naturally follows from the TR transformation combined with  causality.
Moreover, the nonlinear nature of the TR transformation is shown to be the underlying
reason why the renormalized perturbation theory (i.e., the loop-expansion) does not
preserve the FDR order by order.
By introducing a new set of auxiliary fields to linearize the time-reversal
transformation, ABL have attempted to develop a FDR-preserving field theory.
Although ABL's work is a remarkable step forward,
the one-loop results of ABL's theory are found to
have some pathological features. The equation for the nonergodicity parameter
gives nontrivial results even for noninteracting Brownian systems, which
is suspicious and should be examined carefully.
Furthermore, their vertex is ill-behaved, leading to the divergence of the
memory integral.
We tend to believe that these ill-behaved results
are intimately connected to their linearization scheme.

In this Communication, by proposing a simpler but crucial linearization scheme,
we show that the one-loop result in the FDR-preserving
renormalized perturbation theory yields a closed dynamic equation for
the density correlation function, and show that
this closed equation turns out to be the same as that in the standard MCT.
We thus have established the precise relationship of the FDR-preserving 
field theory with the standard MCT.

We start with the following Langevin equation for the density field
$\rho({\bf r},t)$ of interacting Brownian particles
\be
\partial_t \rho({\bf r},t)=
\nabla \cdot \Big( \rho({\bf r},t) \nabla \frac{\delta F[\rho]}
{\delta \rho({\bf r},t)} \Big) +\eta({\bf r},t)
\la{eqn:k1}
\ee
where the Gaussian thermal noise $\eta({\bf r},t)$
has zero mean and variance of the form
\be
<\eta ({\bf r},t)\eta ({\bf r}',t')> = 2T \nabla \cdot \nabla' \Big(
 \rho({\bf r},t)  \delta ({\bf r}-{\bf r}')\delta (t-t')\Big)
 \la{eqn:k2}
\ee
Note that the noise correlation depends on the density variable, i.e.,
the noise is multiplicative.
In (\ref{eqn:k1}), $F[\rho]$ is the free energy density functional
which takes the following form:
\be
F[\rho]= T \int d{\bf r} \,
 \rho ({\bf r}) \Big(\ln \frac{\rho ({\bf r})}{\rho_0}-1 \Big) 
+ \frac{1}{2} \int  d {\bf r} \int  d {\bf r}'
\, \delta \rho ({\bf r})\, U({\bf r}-{\bf r}')\,  \delta \rho ({\bf r}')
\la{eqn:k3}
\ee
where $\delta \rho ({\bf r},t) \equiv \rho({\bf r},t)-\rho_0$
is the density fluctuation around the equilibrium density $\rho_0$.
In (\ref{eqn:k3}) the first term is the ideal gas part of the free energy,
$F_{id}[\rho]$, and the second term the interaction part of the free energy,
$F_{int}[\rho]$.
Using Ito calculus, Dean \cite{dean} has derived
the above nonlinear Langevin equation for
the density field of system of interacting Brownian particles with pair potential
$ U({\bf r})$. Earlier, Kawasaki \cite{kkddft} has obtained
the same Langevin equation with $U({\bf r})$ replaced by
$-T c ({\bf r}),\,\, c({\bf r})$ being the direct correlation function,
by adiabatically eliminating the momentum field in the 
fluctuating nonlinear hydrodynamic equations \cite{dasmazenko}
of the glass-forming liquids.
For this case, the free energy density functional (\ref{eqn:k3})
 takes the Ramakrishnan-Yussouff form.

We consider the corresponding action integral ${\cal S}[\rho, \hat \rho]$
which governs the stochastic dynamics of the coarse grained density variable
\be
{\cal S}[\rho, \hat \rho] = \int \, d {\bf r} \int dt \,
 \Big\{ \hat \rho \Big[ -\partial_t \rho
+\nabla \cdot \Big( \rho \nabla \frac{\delta F}{\delta \rho} \Big)  \Big]
 +T \rho ( \nabla \hat \rho)^2 \Big\}
\la{eqn:k4}
\ee
where the field $\hat\rho$ is pure imaginary, and the
last term comes from the multiplicative thermal noise.
(A similar action was given in \cite{kkft} but with real  $\hat \rho$.)
It is a crucial observation of ABL to recognize that the above action is
{\em invariant} under the TR field transformation:
\bea
\quad \rho ({\bf r}, -t)& = & \rho ({\bf r}, t) \nonumber \\
\hat \rho ({\bf r}, -t) & = & -\hat \rho ({\bf r}, t)
+ \frac{1}{T} \frac{\delta F}{\delta \rho ({\bf r},t)}
\la{eqn:k5}
\eea

The FDR follows from the above TR transformation.
The response function $R({\bf r},t; {\bf r}' t')$ is defined as a link between
induced density change $\Delta <\rho({\bf r},t)>$ and
an external  infinitesimal field  $h_e({\bf r}',t')$:
\be
\Delta <\rho( {\bf r},t)> \equiv \int d {\bf r}' \,
\int dt'  \,  R({\bf r},t; {\bf r}' t') \,  h_e({\bf r}',t')
\la{eqn:k6}
\ee
where $< \cdots > \equiv \int {\cal D}\rho \int {\cal D}{\hat \rho} \, (\cdots)
\exp({\cal S}[\rho, \hat\rho])$.
To find the induced change of density, one should add
the contribution of the external field to the free energy $F$,
$\Delta F \equiv -\int d{\bf r} \int dt \,
\delta \rho({\bf r},t) h_e({\bf r},t)$.
It is straightforward to show, by obtaining the induced density change, that
the response function $R( {\bf r},t; {\bf r}' t')$ is given by
\be
R({\bf r},t; {\bf r}' t') =-\Big<  \rho( {\bf r},t) \nabla' \cdot \Big( \rho( {\bf r}',t')
\nabla' {\hat \rho}( {\bf r}',t') \Big) \Big>
\la{eqn:k7}
\ee
Note that the response function  is {\em not} given by
the conventional response function 
$-<\rho( {\bf r},t))\nabla'^2{\hat \rho}( {\bf r}',t')>$ which is
the true response function for the Langevin equation with additive noise.
Instead, the response function is given by (\ref{eqn:k7})
 due to the multiplicative nature of the Langevin equation  noise
for the density fluctuation \cite{miya,abl}.
Using the identity $\Big< \rho ({\bf r},t) \delta {\cal S}/ 
\delta {\hat \rho}({\bf r}',t') \Big>=0$ and the TR transformation 
(\ref{eqn:k5}),  one obtains the FDR
\be
-\frac{1}{T} \partial_{t} G_{\rho \rho}({\bf r}-{\bf r}', t-t')
=-R({\bf r}-{\bf r}', t'-t) + R({\bf r}-{\bf r}', t-t')
\la{eqn:k8}
\ee
where $G_{\rho \rho}({\bf r}-{\bf r}', t-t') \equiv
\big< \delta \rho ({\bf r},t) \delta \rho ({\bf r}',t')  \big> $.
Since $ R({\bf r}-{\bf r}', t'-t)=0$ for $t > t'$ due to causality,
(\ref{eqn:k8}) gives the standard form of the FDR
\be
R({\bf r}-{\bf r}', t-t')=-\Theta(t-t')\frac{1}{T}\partial_t
G_{\rho \rho}({\bf r}-{\bf r}', t-t')
\la{eqn:k9}
\ee
where $\Theta(t)$ is the Heaviside step function.

Using the form of the free energy given in (\ref{eqn:k3}),
one can explicitly write $\delta F/\delta \rho$  as
\bea
\frac{1}{T} \frac{\delta F_{id}[\rho]}{\delta \rho ({\bf r},t)}
 &=& \ln \frac{\rho ({\bf r},t)}{\rho_0}
 \equiv \frac{\delta \rho ({\bf r},t)}{\rho_0}+ f(\delta \rho ({\bf r},t)), \nonumber \\
\frac{1}{T} \frac{\delta F_{int}[\rho]}{\delta \rho ({\bf r},t)}
&=& \frac{1}{T} \int \, d {\bf r}'
 U({\bf r}-{\bf r}') \delta \rho ({\bf r}',t)
\la{eqn:k10}
\eea
where   $f[\delta \rho({\bf r},t)]$
is the contribution of the non-Gaussian part of $F_{id}[\rho]$,
 and is given by an infinite series
$f[\delta \rho({\bf r},t)] \equiv -\sum_{n=2}^{\infty}
\frac{1}{n} \big(-\delta \rho ({\bf r},t)/\rho_0 \big)^n$.
 Using (\ref{eqn:k10}) one can explicitly write
the TR transformation (\ref{eqn:k5}) as
\bea
\rho ({\bf r}, -t) &=&  \rho ({\bf r}, t) \nonumber \\
\hat \rho ({\bf r}, -t) &=&  -\hat \rho ({\bf r}, t) 
+f[\delta \rho({\bf r},t)]+{\hat K} \ast \delta \rho({\bf r},t)
\la{eqn:k11}
\eea
where $\hat K*$ is convolution with the kernel
 $K({\bf r})\equiv \big(\delta({\bf r})/\rho_0+U({\bf r})/T \big)$.
Note that (\ref{eqn:k11}) is {\em nonlinear}
due to the non-Gaussian contribution $f[\delta \rho({\bf r},t)]$
to the ideal-gas  part of the free energy \cite{ideal}.
ABL has shown that this nonlinear nature of the transformation
is the origin of the incompatibility of the renormalized perturbation theory
with FDR. Naturally, this incompatibility would be
resolved if the transformation (\ref{eqn:k11}) is made linear 
by ignoring $f[\delta \rho({\bf r},t)]$. 
However, as shown below, in this Gaussianized case,
the nonlinear term  generated by dynamics (the second term
in (\ref{eqn:k13}) below) would give rise to the spurious contribution
to the one-loop result, incorrectly yielding a nontrivial result
\cite{miya,abl,sdd} even in the noninteracting system.

As the most natural way to make the transformation (\ref{eqn:k11}) linear,
we introduce a new field $\theta({\bf r},t)$ so as to satisfy the
nonlinear constraint
\be
\theta ({\bf r},t) = f[\delta \rho({\bf r},t)] \equiv -\sum_{n=2}^{\infty}
\frac{1}{n} \Big(-\frac{\delta \rho}{\rho_0} \Big)^n
\la{eqn:k12}
\ee
Note that our approach here differs from that of ABL in that whereas ABL
defines the new variable as the functional derivative of the full free energy
with respect to density: $\theta_{ABL}({\bf r},t)
 \equiv \delta F/\delta \rho ({\bf r},t)$,
we limit the new variable $\theta({\bf r},t)$ only to the
nonlinear part of it.
 Using the constraint (\ref{eqn:k12}),
we obtain the ideal-gas contribution to  the body force as
\be
\nabla \cdot \Big( \rho \nabla \frac{\delta F_{id}}
{\delta \rho} \Big)
 = T\nabla^2 \rho +  \frac{T}{\rho_0} \nabla \cdot \big( \delta \rho  \nabla \rho \big)
 +\rho_0 T \nabla^2 \theta +T \nabla \cdot \big( \delta \rho  \nabla \theta \big)
\la{eqn:k13}
\ee
where the first (the last) two terms are the contributions from the
Gaussian (non-Gaussian) parts of the ideal-gas free energy.
Since due to cancellation of the two nonlinear effects,
the entire ideal-gas contribution to the dynamics is pure diffusion, i.e.,
$\nabla \cdot \Big( \rho \nabla \delta F_{id}/\delta \rho \Big)=T\nabla^2 \rho$,
 the sum of the last three terms in (\ref{eqn:k13}) should {\em vanish if}
the constraint (\ref{eqn:k12}) is taken into account:
\be
 \frac{T}{\rho_0} \nabla \cdot \big( \delta \rho  \nabla \rho \big)
 +\rho_0 T \nabla^2 \theta +
T \nabla \cdot \big( \delta \rho  \nabla \theta \big)=0
\la{eqn:k14}
\ee
As shown below, this nonperturbative cancellation plays a crucial role in obtaining
a closed equation for the density correlation function alone.

Incorporating the new variable $\theta({\bf r},t)$ and its conjugate ${\hat \theta}({\bf r},t)$,
the action (\ref{eqn:k4}) can now be explicitly rewritten as
\bea
{\cal S}[\psi] &\equiv& {\cal S}_g[\psi] +{\cal S}_{ng}[\psi] \nonumber \\
{\cal S}_g[\psi]  &\equiv&
\int d{\bf r} \int dt \, \Big\{ \hat \rho \Big[ -\partial_t \rho
+T \nabla^2 \rho +\u{\rho_0 T \nabla^2 \theta}
+\rho_0 \nabla^2 {\hat U} \ast \delta \rho \Big]
 +T \rho_0 ( \nabla \hat \rho)^2 +  \hat \theta \theta \Big\} \nonumber \\
{\cal S}_{ng} [\psi]  &\equiv&
\int d{\bf r} \int dt \, \Big\{ \hat \rho \Big[
 \nabla \cdot \Big(\delta \rho \nabla {\hat U} \ast \delta \rho \Big)+
 \u{\frac{T}{\rho_0} \nabla \cdot \big( \delta \rho \nabla \rho \big)}
+ \u{T \nabla \cdot \big( \delta \rho \nabla \theta \big)} \Big] \nonumber \\
&{}&+ \, T \delta \rho ( \nabla \hat \rho)^2 - \hat \theta f(\delta\rho) \Big\}
\la{eqn:k15}
\eea
where $\psi$ denotes the entire set of the fields collectively, and
the full action ${\cal S}[\psi]$ is separated into
its Gaussian part ${\cal S}_g[\psi]$ and non-Gaussian part
${\cal S}_{ng}[\psi]$ \cite{free}.
Now the actions  ${\cal S}_g[\psi]$
and ${\cal S}_{ng}[\psi]$ are {\em separately} invariant
under the following {\em linear} TR transformation
\bea
 \quad \rho ({\bf r}, -t)& = & \rho ({\bf r}, t) \nonumber \\
\hat \rho ({\bf r}, -t) & = & -\hat \rho ({\bf r}, t) +
 \theta({\bf r},t) + {\hat K} \ast \delta \rho({\bf r},t)  \nonumber \\
\theta ({\bf r}, -t) & = & \theta ({\bf r}, t) \nonumber \\
\hat \theta ({\bf r}, -t) & = & \hat \theta ({\bf r}, t)
- \partial_t \rho ({\bf r},t)
\la{eqn:k16}
\eea
It is easy to show that  the modulus of the associated transformation matrix is unity.
Though the three underlined terms in (\ref{eqn:k15})
vanish when summed together, their presence is
crucial for the actions ${\cal S}_g[\psi]$ and ${\cal S}_{ng}[\psi]$
to be separately time-reversal invariant.
Note also that the separate invariance of the actions under
the linear transformation (\ref{eqn:k16}) is not tied to
the form of the constraint (\ref{eqn:k12}).
This separate invariance of  ${\cal S}_g[\psi]$
and ${\cal S}_{ng}[\psi]$ enables us to construct the FDR-preserving renormalized
perturbation theory from these actions.

 
With the new action (\ref{eqn:k15}), it is easy to show that  
the response function is given by
\be
R({\bf r},t; {\bf r}',t')=\frac{1}{T} \Big< \delta \rho( {\bf r},t)\,
 {\hat \theta}( {\bf r}',t') \Big>
 \la{eqn:k17}
\ee
One can  then obtain the FDR (\ref{eqn:k8})
by taking correlation of the last member of (\ref{eqn:k16}) 
with $\delta \rho({\bf r},t)/T$.

Incorporating the above new set of variables into the theory,
we are now ready to develop a renormalized perturbation theory 
which preserves the FDR order by order.
 We first consider the noninteracting case ($U=0$) and show
that the noninteracting part of the action,
${\cal S}_{id}[\psi] \equiv {\cal S}[\psi;  U=0]$,
yields the dynamic behavior consistent for the noninteracting system.
We begin with the following identities
\be
\Big<\delta \rho({\bf 2}) \frac{\delta {\cal S}_{id}[\psi]}
{\delta {\hat \rho} ({\bf 1})} \Big> =0,
\qquad
\Big< \delta \rho({\bf 2}) \frac{\delta {\cal S}_{id}[\psi]}
{\delta  \theta ({\bf 1})} \Big> =0
\la{eqn:k18}
\ee
where ${\bf 1} \equiv ({\bf r}, t)$ and ${\bf 2} \equiv ({\bf 0}, 0)$.
The first identity can be written explicitly as
\be
0=\Big<\delta \rho({\bf 2}) \frac{\delta {\cal S}_{id}[\psi]}
{\delta \hat \rho ({\bf 1})} \Big>
= \Big(-\frac{\partial}{\partial t} + T \nabla^2  \Big)G_{\rho \rho}
({\bf1}-{\bf 2}) -2T\rho_0 \nabla^2 \big< \hat \rho({\bf 1})
 \delta \rho({\bf 2}) \big>
-2T \big< \delta \rho ({\bf 2}) \nabla \cdot \big(\delta \rho ({\bf 1}) \nabla
\hat \rho ({\bf 1}) \big)\big>
\la{eqn:k19}
\ee
where we used the fact that the sum of the three underlined terms in
(\ref{eqn:k15}) vanishes.
Similarly, using the second identity in (\ref{eqn:k18}),
we obtain
\be
0=\Big<\delta \rho({\bf 2}) \frac{\delta
{\cal S}_{id}[\psi]}{\delta \theta ({\bf 1})} \Big>
=\rho_0 T \nabla^2 \big< \hat \rho ({\bf 1}) \delta \rho({\bf 2}) \big>
+\big<\hat \theta({\bf 1}) \delta \rho({\bf 2}) \big>
+T \big< \delta \rho ({\bf 2}) \nabla \cdot \big(\delta \rho ({\bf 1}) \nabla
\hat \rho ({\bf 1}) \big)\big>
\la{eqn:k20}
\ee
In (\ref{eqn:k20}), $\big< \hat \rho ({\bf 1}) \delta \rho({\bf 2}) \big>=0$ an
$\big<\hat \theta({\bf 1}) \delta \rho({\bf 2}) \big>=0$ for $t > 0$ due to
the causality, and hence we obtain
\be
\big< \delta \rho ({\bf 2}) \nabla \cdot \big(\delta \rho ({\bf 1}) \nabla
\hat \rho ({\bf 1}) \big)\big>=0   \quad \mbox{for} \quad t >0
\la{eqn:k21}
\ee
Using (\ref{eqn:k21}) and causality,  we obtain from (\ref{eqn:k19})
\be
\frac{\partial}{\partial t}G_{\rho \rho}({\bf r}, t)
 = T \nabla^2  G_{\rho \rho}({\bf r}, t) \quad \mbox{for} \quad t >0
\la{eqn:k22}
\ee
This result is  expected  for the non-interacting system.

There are  five nonlinear terms in the full action (\ref{eqn:k15}) 
where the two underlined nonlinear terms 
$(T/\rho_0)\hat\rho \nabla \cdot \Big( \delta \rho \nabla \rho \Big)$
and $T \hat\rho \nabla \cdot \Big( \delta \rho \nabla \theta \Big)$
are shown to cancel the linear diffusion term
${\hat \rho} \rho_0 T \nabla^2 \theta$.
Then the remaining nonlinearities are
$\hat\rho \nabla \cdot \Big( \delta \rho \nabla {\hat U}\ast \delta \rho
\Big)$, $T \delta \rho \big(\nabla \hat\rho \big)^2$, and
$\hat\theta f[\delta \rho]$.
The first two come from the particle interaction and
 the multiplicative thermal noise, respectively.
The last cubic nonlinear term
$\frac{1}{2}\hat\theta (\delta \rho/\rho_0)^2$,
the only one contributing in the one loop order,
comes from the non-Gaussian part of the ideal-gas free energy.
In order to analyze the effect of these three nonlinear terms,
one should examine the structures of the relevant self-energies.

Let us write
\bea
 {\cal S}_g[\psi] &=&
 \frac{1}{2}\psi^T({\bf 1})\cdot G_0^{-1}({\bf 12})\cdot\psi({\bf 2})\nonumber\\{\cal S}_{ng} [\psi]  &=&  \frac{1}{3!}V(123)\psi({ 1})
\psi({ 2})\psi({3})+
\hat\theta({\bf 1})\sum_{n=3}^{\infty}\frac{1}{n}
\Big(-\frac{\delta \rho({\bf 1})}{\rho_0}\Big)^n
\la{eqn:k23}
\eea
where $\psi({\bf 1})$ and $\psi^T({\bf 1})$ are respectively
column and row vectors of the
four fields $\rho$, $\hat\rho$, $\theta$, and $\hat\theta$,
 and the term with $n=2$ in the summation is incorporated into $V(123)$.
The unperturbed inverse matrix propagator $G^{-1}_0({\bf 12})$,
 can be read off from the action (\ref{eqn:k15}) as
\be
G_0^{-1}({\bf 12})= \left(\begin{array}{rrrr}
0\qquad\qquad&{\tilde D}_1\delta({\bf 12})+\rho_0\nabla_1^2
  U({\bf 12})& 0\qquad\qquad&0\\
D_1\delta({\bf 12})+\rho_0\nabla_1^2  U({\bf 12})&
-2T\rho_0\nabla_1^2\delta({\bf 12})& T\rho_0\nabla_1^2\delta({\bf 12})&0\\
0\qquad\qquad& T\rho_0\nabla_1^2\delta({\bf 12})&
0\qquad\qquad&\delta({\bf 12})\\
0\qquad\qquad&0\qquad\qquad&\delta({\bf 12})\qquad\qquad&0
\end{array}\right)
\la{eqn:k24}
\ee
where $D_1 \equiv (-\partial/\partial t_1 + T \nabla^2_1)$ and 
${\tilde D}_1 \equiv (\partial/\partial t_1 + T \nabla^2_1)$.
$V(123)$ in fact is a collection of vertices
$V_{\al_1\al_2\al_3}({\bf 123})$ with the $\al$'s standing for four field types,
  $\bf 1,2,3\cdots$ for space-time coordinates,
and $\psi(1)$ etc stand for $\psi_{\al_1}({\bf 1})$ etc.
Repeated thin (thick) numbers indicate space-time integrations and summations
over the field types (space-time integrations only).
We give expressions for the vertices $V_{\al_1\al_2\al_3}({\bf 123})$:
\bea
V^{id}_{\hat\rho\rho\rho}({\bf 123})&=&-\frac{T}{\rho_0}
\Big[{\BF\nabla}_1 \delta({\bf 12}) \cdot {\BF\nabla}_3 \delta ({\bf 23})
+{\BF\nabla}_1 \delta({\bf 13}) \cdot {\BF\nabla}_2 \delta ({\bf 23}) \Big],
\nonumber \\
V^{int}_{\hat\rho\rho\rho}({\bf 123})&=&
-\Big[ {\BF \nabla}_1 \delta ({\bf 12}) \cdot {\BF \nabla}_3 {\hat U}({\bf 23})
+{\BF \nabla}_1 \delta ({\bf 13}) \cdot {\BF \nabla}_2 {\hat U}({\bf 23})
\Big] \delta(t_2-t_3), \nonumber\\
V_{\hat\rho\rho\theta}({\bf 123})&=&
-T{\BF\nabla}_1\cdot {\BF\nabla}_3\delta ({\bf 123}), \quad
V_{\rho \hat\rho \hat\rho}({\bf 123})=
T{\BF\nabla}_2 \delta ({\bf 12}) \cdot {\BF\nabla}_3 \delta ({\bf 13}),  \nonumber \\
V_{\hat\theta\rho\rho}({\bf 123})&=&
\frac{1}{2\rho_0^2}\delta ({\bf 12})\delta ({\bf 23})
\la{eqn:k25}
\eea
where we have separated the vertex $V_{\hat\rho\rho\rho}({\bf 123})$ into
the ideal-gas contribution $V^{id}_{\hat\rho\rho\rho}({\bf 123})$ and
the interaction contribution $V^{int}_{\hat\rho\rho\rho}({\bf 123})$.
In (\ref{eqn:k24}) and (\ref{eqn:k25}) the space-time delta function is
defined as $\delta ({\bf 12})\equiv \delta ({\bf r}_1-{\bf r}_2)\delta (t_1-t_2)$.

The dynamic equations for the correlation and
response functions are formally given
by the matrix Schwinger-Dyson (SD) equation
\be
G_0^{-1}({\bf 13})\cdot G({\bf 32})  -
 \Sigma ({\bf 13}) \cdot G({\bf 32}) = \delta ({\bf 12})
\la{eqn:k26}
\ee
In order to obtain the equation of motion for the density correlation
function, we compute $\hat\rho \rho$-element of the SD equation by
taking ${\bf 1}\equiv ({\bf r},t)$, ${\bf 2} \equiv ({\bf 0},0)$, and
${\bf 3}\equiv ({\bf r}_s,s)$. 
It is now convenient to introduce the Fourier transform in space
\be
\Sigma({\bf r},t)=\int_{{\bf k}} \Sigma({\bf k},t) e^{i{\bf k}\cdot {\bf r}}
\la{eqn:k27}
\ee
etc. where $\int_{{\bf k}} \equiv \int d {\bf k}/(2\pi)^3$.
Using (\ref{eqn:k24}) and causality of the response functions,
we obtain for $t>0$
\be
 [G^{-1}_0 \cdot G]_{\hat\rho \rho}({\bf k},t)
  = -\Big(\partial_t +\rho_0 T {\bf k}^2  K({\bf k}) \Big) G_{\rho \rho}({\bf k},t)
- \rho_0 T {\bf k}^2 G_{\theta \rho}({\bf k},t)
\la{eqn:k28}
\ee
Likewise using the causalities of  the self-energy functions and
the response functions, we obtain for
$ t>0 $
\bea
\big[\Sigma \cdot G \big]_{\hat\rho \rho}({\bf k},t)
&=& \int_{-\infty}^t ds \,\,
 \Big[ \Sigma_{\hat\rho \rho}({\bf k}, t-s)
G_{\rho \rho}({\bf k},s)
+  \Sigma_{\hat\rho \theta}({\bf k}, t-s)
G_{\theta \rho}({\bf k},s) \Big] \nonumber \\
&+&  \int_{-\infty}^0 ds \,\,
 \Big[ \Sigma_{\hat\rho \hat\rho}({\bf k}, t-s)
G_{\hat\rho \rho}({\bf k},s)
+  \Sigma_{\hat\rho \hat\theta}({\bf k}, t-s)
G_{\hat\theta \rho}({\bf k},s)\Big]
\la{eqn:k29}
\eea
where the upper limits of the time integration are due to
the causality of the self-energy functions in the first
two integrations, and to the causality of the response functions
 in the last two integrations.
It can be  shown that (\ref{eqn:k29}) can be simplified using the TR properties
of the self-energies and the response functions as
\bea
\big[\Sigma \cdot G \big]_{\hat\rho \rho}({\bf k},t)
 = &-& \int_0^t ds \,\,
\Big[ \Sigma_{\hat\rho \hat\theta}({\bf k}, t-s)
\partial_s G_{\rho \rho}({\bf k},s)
+ K({\bf k})  \Sigma_{\hat\rho \hat\rho}({\bf k}, t-s)
 G_{\rho \rho}({\bf k},s) \Big] \nonumber \\
&-& \int_0^t ds \,\,  \Sigma_{\hat\rho \hat\rho}({\bf k}, t-s)
 G_{\theta \rho}({\bf k},s)
\la{eqn:k30}
\eea
Using (\ref{eqn:k28}) and (\ref{eqn:k30}),
the $\hat\rho \rho$-element of the SD equation for $t>0$ can now
be explicitly written as
\bea
\partial_t G_{\rho \rho}({\bf k},t) 
&=& -\rho_0 T {\bf k}^2 K({\bf k}) G_{\rho \rho}({\bf k},t)
-\rho_0 T {\bf k}^2 G_{\theta \rho}({\bf k},t) \nonumber \\
&+& \int_0^t ds \,\,
\Big[ \Sigma_{\hat\rho \hat\theta}({\bf k}, t-s)
\partial_s G_{\rho \rho}({\bf k},s)
+  K({\bf k})  \Sigma_{\hat\rho \hat\rho}({\bf k}, t-s)
 G_{\rho \rho}({\bf k},s) \Big] \nonumber \\
 &+& \int_0^t ds \,\,
 \Sigma_{\hat\rho \hat\rho}({\bf k}, t-s) G_{\theta \rho}({\bf k},s)
\la{eqn:k31}
\eea
One can likewise obtain $\hat\theta \rho$-element of the SD equation as
\bea
 G_{\theta \rho}({\bf k},t)&=& \Sigma_{\hat\theta \hat\theta}({\bf k},0)
G_{\rho \rho}({\bf k},t)
-\int_0^t ds \, \Sigma_{\hat\theta \hat\theta}({\bf k}, t-s)
\partial_s G_{\rho \rho}({\bf k},s) \nonumber \\
&-& \int_0^t ds \,
 \Big[ K({\bf k}) \Sigma_{\hat\theta \hat\rho}({\bf k}, t-s)
G_{\rho \rho}({\bf k},s)
+  \Sigma_{\hat\theta \hat\rho}({\bf k}, t-s)
G_{\theta \rho}({\bf k},s) \Big]
\la{eqn:k32}
\eea

We now take the two crucial steps which can give
a closed equation for $G_{\rho \rho}({\bf r},t)$ in {\em one-loop} order:
\begin{itemize}
\item
The equations (\ref{eqn:k31}) and (\ref{eqn:k32}) imply that
the integral involving $G_{\theta \rho}({\bf r},t)$ in (\ref{eqn:k31})
are of higher order and hence can be discarded
 in the one-loop order (note that $[G_0]_{\theta \rho} =0$ from (\ref{eqn:k24})).
 \item
As we alluded earlier,
using  the cancellation of the three underlined terms in (\ref{eqn:k15}),
   one can eliminate $ -\rho_0 T {\bf k}^2 G_{\theta \rho}({\bf k},t)$ 
in (\ref{eqn:k31}), together with the two vertices
$V^{id}_{\hat\rho \rho \rho}$ and $V_{\hat\rho \rho \theta}$ 
appearing in (\ref{eqn:k30}). 
It should be emphasized that the cancellation is a nonperturbative
effect which is required to hold only when the constraint (\ref{eqn:k12}) is
employed. Therefore,  we discard $-\rho_0 T {\bf k}^2 G_{\theta\rho}({\bf k},t)$
 and retain  only those terms arising from   the remaining three vertices
$V^{int}_{\hat\rho \rho \rho}$, $V_{\rho \hat\rho \hat\rho}$, and
$V_{\hat\theta \rho \rho}$ in evaluating the one-loop self-energies.
\end{itemize}
Taking these steps we obtain the equation of motion
for the density correlation function up to the {\em one-loop order} as
\bea
&{}&\partial_t G_{\rho \rho}({\bf k},t) = 
-\rho_0 T {\bf k}^2 K({\bf k}) G_{\rho \rho}({\bf k},t)
\nonumber \\
&+& \int_0^t ds \,\,
 \Big[ \Sigma_{\hat\rho \hat\theta}({\bf k}, t-s)
\partial_s G_{\rho \rho}({\bf k},s)
 +  K({\bf k})  \Sigma_{\hat\rho \hat\rho}({\bf k}, t-s)
 G_{\rho \rho}({\bf k},s) \Big]
 \la{eqn:k33}
\eea
In the absence of particle interaction ($U=0$),
the equation (\ref{eqn:k33}) reduces to the diffusion equation
(\ref{eqn:k22}) since, as shown below,
the self-energies in (\ref{eqn:k33}) involves
the vertex $V^{int}_{\hat\rho \rho \rho}$ which contains
the interaction potential ${\hat U}$.

Now we compute the one-loop self-energies
$\Sigma_{\hat\rho \hat \theta}({\bf k},t)$ and
$\Sigma_{\hat\rho \hat \rho}({\bf k},t)$.
With the three surviving vertices
$V^{int}_{\hat\rho \rho \rho}$, $V_{\rho \hat\rho \hat\rho}$, and
$V_{\hat\theta \rho \rho}$, we find that
there is the only one nonvanishing
diagram for $\Sigma_{\hat\rho \hat\theta}({\bf 12})$, which is given by
\be
\Sigma_{\hat\rho \hat\theta}({\bf 12})
=  V^{int}_{\hat\rho \rho \rho}({\bf 134}) V_{\hat\theta \rho \rho}({\bf 256})
G_{\rho \rho}({\bf 35}) G_{\rho \rho}({\bf 46})
\la{eqn:k34}
\ee
Likewise, there are three nonvanishing contributions to the self-energy
$ \Sigma_{\hat\rho \hat\rho}({\bf 12})$:
\bea
\Sigma_{\hat\rho \hat\rho}({\bf 12}) &=& \Sigma^{(1)}_{\hat\rho \hat\rho}({\bf 12})
+ \Sigma^{(2)}_{\hat\rho \hat\rho}({\bf 12})  \nonumber \\
\Sigma^{(1)}_{\hat\rho \hat\rho}({\bf 12}) &=&
 V^{int}_{\hat\rho \rho \rho}({\bf 134})
 V^{int}_{\hat\rho \rho \rho}({\bf 256})
G_{\rho \rho}({\bf 35}) G_{\rho \rho}({\bf 46}) \nonumber \\
\Sigma^{(2)}_{\hat\rho \hat\rho}({\bf 12}) &=&
 V^{int}_{\hat\rho \rho \rho}({\bf 134})
V_{\rho \hat\rho \hat\rho}({\bf 562})
G_{\rho \rho}({\bf 35}) G_{\rho \hat\rho}({\bf 46})
\la{eqn:k35}
\eea
$\Sigma^{(1)}_{\hat\rho \hat\rho}$ in (\ref{eqn:k35}) comes solely from the interaction
contribution to the body force, and 
$\Sigma^{(2)}_{\hat\rho \hat\rho}$ 
from the cross-contribution of the interaction and multiplicative thermal noise.
By multiplying the second equation in the transformation (\ref{eqn:k16}) by
$\delta \rho({\bf 0},0)$, taking average and using causality,
one obtains
\be
G_{\rho\hat\rho}({\bf r},t)=\Theta (t)\Big(
{\hat K}\ast G_{\rho\rho}({\bf r},t)+G_{\rho\theta}({\bf r},t)\Big)
\la{eqn:k36}
\ee
When (\ref{eqn:k36}) is substituted into (\ref{eqn:k35}),
the correlation function $G_{\rho\theta}({\bf r},t)$ will make null contribution
in the one-loop order.
Using this fact one can rewrite (\ref{eqn:k35}) as
\bea
\Sigma^{(1)}_{\hat\rho \hat\rho}({\bf 12}) &=&
  V^{int}_{\hat\rho \rho \rho}({\bf 134})
 V^{int}_{\hat\rho \rho \rho}({\bf 256})
G_{\rho \rho}({\bf 35}) G_{\rho \rho}({\bf 46}) \nonumber \\
\Sigma^{(2)}_{\hat\rho \hat\rho}({\bf 12}) &=&
  V^{int}_{\hat\rho \rho \rho}({\bf 134})
V_{\rho \hat\rho \hat\rho}({\bf 562})
 G_{\rho \rho}({\bf 35})  {\hat K}\ast G_{\rho\rho}({\bf 46}) \Theta(t_4-t_6)
\la{eqn:k37}
\eea
Therefore the equations (\ref{eqn:k33})-(\ref{eqn:k34}) and (\ref{eqn:k37})
consist of a closed equation for
the density correlation function $G_{\rho \rho}({\bf r},t)$ alone.


The Fourier components of the self-energies are now computed as
\bea
\Sigma_{\hat\rho \hat\theta}({\bf k},t)&=&-\frac{1}{2\rho_0^2} \int_{{\bf q}}
V({\bf k},{\bf q}) G_{\rho \rho}({\bf q},t)
G_{\rho \rho}({\bf k}-{\bf q},t) \nonumber \\
\Sigma^{(1)}_{\hat\rho \hat\rho}({\bf k},t)
&=&  \int_{{\bf q}} V^2({\bf k},{\bf q}) 
G_{\rho \rho}({\bf q},t) G_{\rho \rho}({\bf k}-{\bf q},t) \nonumber \\
\Sigma^{(2)}_{\hat\rho \hat\rho}({\bf k},t)
&=& -\frac{1}{2} \int_{{\bf q}}
V^2({\bf k},{\bf q})
 G_{\rho \rho}({\bf q},t) G_{\rho \rho}({\bf k}-{\bf q},t) \nonumber \\
&-& \frac{T}{2\rho_0} {\bf k}^2 \int_{{\bf q}}
V({\bf k},{\bf q})G_{\rho \rho}({\bf q},t) G_{\rho \rho}({\bf k}-{\bf q},t)
\la{eqn:k38}
\eea
where $V({\bf k},{\bf q})\equiv  \big[( {\bf k}\cdot{\bf q}) U({\bf q})
+{\bf k}\cdot({\bf k}-{\bf q}) U({\bf k}-{\bf q}) \big]$.
Arranging the sum of the self-energies
$\Sigma^{(i)}_{\hat\rho \hat\rho}$ with $i=1,2$,
one can rewrite $\Sigma_{\hat\rho \hat\rho}({\bf k},t)$ as
\bea
\Sigma_{\hat\rho \hat\rho}({\bf k},t) &\equiv& {\bf k}^2 \,
{\tilde \Sigma}_{\hat\rho \hat\rho}({\bf k},t) \nonumber \\
{\tilde \Sigma}_{\hat\rho \hat\rho}({\bf k},t) &=&
\frac{1}{2} \int_{{\bf q}} 
\Big[{\hat V}^2({\bf k},{\bf q})-\frac{T}{\rho_0}V({\bf k},{\bf q})\Big]
G_{\rho \rho}({\bf q},t) G_{\rho \rho}({\bf k}-{\bf q},t) 
\la{eqn:k39}
\eea
with ${\hat V}({\bf k},{\bf q})\equiv V({\bf k},{\bf q})/|{\bf k}|$.

In order to appreciate what the closed equation (\ref{eqn:k33}) implies,
let us for the moment ignore the contribution of the self-energy
$\Sigma_{\hat\rho \hat\theta}({\bf k},t)$ in (\ref{eqn:k28}).
Then (33) would become
\be
\partial_t G_{\rho \rho}({\bf k},t) = -\rho_0 T {\bf k}^2  K({\bf k})
G_{\rho \rho}({\bf k},t)
+  {\bf k}^2 K({\bf k}) \int_0^t ds \,\,
 {\tilde \Sigma}_{\hat\rho \hat\rho}({\bf k}, t-s)
 G_{\rho \rho}({\bf k},s)
\la{eqn:k40}
\ee
One can see from (\ref{eqn:k40}) that the effect of the nonlinear contribution
${\tilde \Sigma}_{\hat\rho \hat\rho}({\bf k},t)$ is to renormalize
the 'bare' life time $\tau_0({\bf k}) \equiv \big(\rho_0 T {\bf k}^2
 K({\bf k}) \big)^{-1}$\cite{kkft} as follows.
Defining the Laplace transform $ G^L_{\rho \rho}({\bf k},z) \equiv
\int_0^{\infty} dt \, e^{-z t} G_{\rho \rho}({\bf k},t)$, etc.,
we obtain the Laplace transform of (\ref{eqn:k40}) as
\be
G^L_{\rho \rho}({\bf k},z) = G_{\rho \rho}({\bf k},0) \cdot
\Big[z+ \tau^{-1}_R ({\bf k},z)    \Big]^{-1}
\la{eqn:k41}
\ee
where we defined the renormalized life time as
\be
\tau_R({\bf k},z) \equiv \tau_0({\bf k})
\Big(1-\frac{1}{\rho_0 T}
{\tilde \Sigma}^L_{\hat\rho \hat\rho}({\bf k}, z)\Big)^{-1}
  =\tau_0({\bf k}) \Big(1+ \frac{1}{\rho_0 T}{\tilde
  \Sigma}^L_{\hat\rho \hat\rho}({\bf k}, z)\Big),
\la{eqn:k42}
\ee
the last equality holding in the one-loop theory.
Substituting (\ref{eqn:k42}) into (\ref{eqn:k41}),
one obtains
\be
G^L_{\rho \rho}({\bf k},z) = G_{\rho \rho}({\bf k},0) \cdot
\Big[ z+ \frac{\tau^{-1}_0({\bf k})}{1+{\tilde
\Sigma}^L_{\hat\rho \hat\rho}({\bf k}, z)/\rho_0 T}    \Big]^{-1}
\la{eqn:k43}
\ee
The corresponding equation in time domain is given by
\be
\partial_t G_{\rho \rho}({\bf k},t) = -\rho_0 T {\bf k}^2  K({\bf k})
G_{\rho \rho}({\bf k},t)
+   \int_0^t ds \,\,
 {\tilde \Sigma}_{\hat\rho \hat\rho}({\bf k}, t-s)
  \Big(- \partial_s G_{\rho \rho}({\bf k},s) /\rho_0 T \Big)
\la{eqn:k44}
\ee
The equation (\ref{eqn:k44}) is in fact obtained by replacing
$G_{\rho \rho}({\bf k},s)$ in the convolution integral  in (\ref{eqn:k40}) 
by $\big(-\partial_s G_{\rho \rho}({\bf k},s)/\rho_0 T {\bf k}^2 K({\bf k})\big)$,
which is valid up to the one-loop order.

 It is clear that the so far ignored self-energy
$\Sigma_{\hat\rho \hat\theta}({\bf k},t)$
 makes a new contribution to the renormalization of the bare life time
 $\tau_0({\bf k})$, which comes from the consistency requirement of the perturbation
theory with the FDR.
Now restoring the contribution of
$\Sigma_{\hat\rho \hat\theta}({\bf k},t)$ and adding it  to (\ref{eqn:k44}),
we obtain
 \bea
\partial_t G_{\rho \rho}({\bf k},t) &=& -\rho_0 T {\bf k}^2  K({\bf k})
G_{\rho \rho}({\bf k},t) -   \int_0^t ds \,\,
  \Sigma_{MC}({\bf k}, t-s) \partial_s G_{\rho \rho}({\bf k},s), \nonumber \\
   \Sigma_{MC}({\bf k},t) &\equiv&
\Big(-\Sigma_{\hat\rho \hat\theta}({\bf k},t)+
\frac{1}{\rho_0 T} {\tilde \Sigma}_{\hat\rho \hat\rho}({\bf k},t) \Big) \nonumber \\ 
&=& \frac{1}{2\rho_0T} \int_{{\bf q}}
{\hat V}^2({\bf k},{\bf q}) G_{\rho \rho}({\bf q},t) G_{\rho \rho}({\bf k}-{\bf q},t)
\la{eqn:k45}
\eea
This equation reduces to the closed dynamic equation for the density correlation
function in the standard MCT if $ U({\bf k})$ is replaced by
$-Tc({\bf k})$. The self-energy $\Sigma_{\hat\rho \hat\theta}({\bf k},t)$
and the part of the self-energy
$\Sigma^{(2)}_{\hat\rho \hat\rho}({\bf k},t)$ which is multiplied by
$\rho_0T$  cancel against each other to yield the last line in (\ref{eqn:k45}).

The equation for the non-ergodicity parameter (NEP) $f({\bf k})$ defined as
$f({\bf k}) \equiv G_{\rho \rho}({\bf k},t 
\rightarrow \infty)/G_{\rho \rho}({\bf k},0)$,
can be obtained from (\ref{eqn:k45}) as
\be
\frac{f({\bf k})}{1-f({\bf k})}=\frac{\Sigma_{MC}({\bf k},\infty)}
{\rho_0 T {\bf k}^2 K({\bf k})}
=\frac{1}{2\rho_0 T^2{\bf k}^2}\int_{{\bf q}} {\hat V}^2({\bf k},{\bf q})
  S({\bf k}) S({\bf q})S({\bf k}-{\bf q}) f({\bf q}) f({\bf k}-{\bf q})
\la{eqn:k46}
\ee
In obtaining (\ref{eqn:k46}), we used
the inverse relationship between $K({\bf k})$ and the static
structure factor $S({\bf k})$: $\rho_0  K({\bf k})=
 \big(1+\rho_0 U({\bf k})/T\big)=\big(1-\rho_0 c({\bf k})\big)=S^{-1}({\bf k})$.
 The NEP equation (\ref{eqn:k46}) is similar to the corresponding one in ABL,
except that there is no memory kernel involved in ABL.
It naturally occurs  from (\ref{eqn:k46}) that $f({\bf k})=0$ is the only
solution for the noninteracting case ($U=0$).
But this feature is absent in ABL's theory.

In summary, in order to obtain the FDR-preserving perturbation theory,
we here proposed a simpler but crucial linearization scheme
in which the new variable is employed to take care
of only the nonlinear part of the TR transformation.
We then recognized that there is a characteristic
nonperturbative cancellation effect in the theory.
This feature enables us to obtain  in the one-loop theory
a closed dynamic equation for the two-point density correlation function.
This equation is shown to be the same as that of the standard MCT.

Having established the relation of our field-theoretical
treatment with the standard MCT, we are now at
the starting point to venture into ambitious tasks
such as higher-order loop calculations and multibody correlations.

\acknowledgements
BK was supported by Grant R01-2003-000-11595-0 from the Basic
Research Program of the Korea Science and Engineering Foundation.
KK was supported by Grant-in-Aid for Scientific Research Grant
17540366 by Japan Society of Promotion of Science.
The final form of this paper was completed while both authors
stayed at the International Institute for Advanced Studies.


\begin{thebibliography}{99}
\bibitem{ideal} Throughout the paper
the term ''ideal gas" is meant only for static behavior.
Dynamics for our system is of course different from the ideal gas
behavior even without interparticle interactions.
\bibitem{free}
 We remind that separation into
Gaussian and non-Gaussian parts do not coincide
for the action $\cal S$ and for the free energy functional $F$.
\end{thebibliography}
\end{document}